\documentclass[aps,twocolumn,pra,twoside,amssymb,amsmath]{revtex4}

\usepackage[latin1]{inputenc}
\usepackage{bm}
\usepackage{multirow,amssymb,amsbsy,amsmath}
\usepackage{graphicx}
\usepackage{mathrsfs}
\usepackage{verbatim}
\usepackage{amssymb}
\usepackage{amsfonts}
\usepackage{bbm}
\usepackage{amsmath}
\usepackage{lmodern}
\usepackage{appendix}
\usepackage{color,soul}
\usepackage[colorlinks,citecolor=blue]{hyperref}
\usepackage{ulem}

\newcommand{\idol}{\ensuremath{\mathbbm 1}}

\newtheorem{theorem}{Theorem}
\newcommand{\tr}{{\rm Tr}}

\makeatletter
\usepackage{pifont}
\makeatother

\usepackage{dcolumn}
\usepackage{epstopdf}
\usepackage{subfigure}

\begin{document}

\title{Experimentally accessible lower bounds for genuine multipartite entanglement and coherence measures}
\author{Yue Dai$^{1,2}$}
\author{Yuli Dong$^2$}
\author{Zhenyu Xu$^2$}
\author{Wenlong You$^2$}
\author{Chengjie Zhang$^{1,2,3}$}
\email{chengjie.zhang@gmail.com}
\author{Otfried G\"uhne$^{3}$}
\email{otfried.guehne@uni-siegen.de}
\affiliation{$^1$School of Physical Science and Technology, Ningbo University, Ningbo, 315211, China\\
$^2$School of Physical Science and Technology, Soochow University, Suzhou, 215006, China\\
$^3$Naturwissenschaftlich-Technische Fakult\"at, Universit\"at Siegen, Walter-Flex-Stra{\ss}e 3, 57068 Siegen, Germany}

\begin{abstract}
Experimentally quantifying entanglement and coherence are extremely important for quantum resource
theory. However, because the quantum state tomography requires exponentially growing measurements
with the number of qubits, it is hard to quantify entanglement and coherence based on the full information
of the experimentally realized multipartite states. Fortunately, other methods have been found to directly
measure the fidelity of experimental states without quantum state tomography. Here we present a fidelity-based
method to derive experimentally accessible lower bounds for measures of genuine multipartite
entanglement and coherence. On the one hand, the method works for genuine multipartite entanglement
measures including the convex-roof extended negativity, the concurrence, the G-concurrence, and the
geometric measure for genuine multipartite entanglement. On the other hand, the method also delivers
observable lower bounds for the convex roof of the $l_{1}$-norm of coherence, the geometric measure of coherence,
and the coherence of formation. Furthermore, all the lower bounds are based on the fidelity between
the chosen pure state and the target state, and we obtain the lower bounds of several real experimental
states as examples of our results.
\end{abstract}

\maketitle

\section{Introduction}
Quantum entanglement is one of the most important concept in quantum mechanics.
In recent years, entanglement has been considered as a valuable resource for
quantum information processing and attracted much attention. For quantifying
entanglement, several entanglement measures have been proposed for bipartite
systems \cite{EOFde2,EOFde21,EOFde22,EOFde23,EOFde24}, such as the the negativity or extensions thereof
\cite{ppt,ppt1,negativity,negativity21,CREN}, the concurrence \cite{EOFde1,2qubit1,2qubit2,concurrence3,concurrence4,concurrence5},
the G-concurrence \cite{Gour,Fan,Uhlmann,GME1} and the geometric measure
of entanglement \cite{GME1,GME2,GME21}.

Compared with bipartite systems, the situation for multipartite systems is
much more complex. Considering an $N$-partite system, there exist different
classes of entanglement \cite{k-entanglement1,k-entanglement2,k-entanglement3}.
A multipartite quantum state that is not a convex combination of biseparable
states with respect to any bipartition contains genuinely multipartite entanglement
(GME) \cite{GME3,GME4,GME5}.  It is worth noticing that the GME denotes the
strongest entanglement type in multipartite quantum states, which is considered
as a rich resource for quantum information processing. Consequently, many criteria
have been introduced for multipartite entanglement detection \cite{bounds15,bounds16,GME11}.

In fact, analytical results concerning the computation of entanglement measures have
been obtained only for special measures and for two-qubit states or some special kinds
of higher-dimensional mixed states \cite{2qubit2,eof1,GME2,iso3,werner,CREN,bounds13}.
It has been proved that computing faithful entanglement measures is NP-hard for a general
state \cite{NP,yichen}. For general higher-dimensional states and multipartite states,
lower or upper bounds are usually presented to quantify entanglement  \cite{Horodecki,Audenaert,mintert04,bounds9,bounds10,bounds11,bounds12,bounds14,bounds141,bounds17,bounds171,ob2}.

Coherence is another crucial quantum mechanical phenomenon and was
characterized in quantum optics. While quantum entanglement can only occur
in bipartite or multipartite systems, quantum coherence is usually defined
for a single system \cite{coherence,rev1,rev2,rev3}. Recently, it has been
recognized that coherence, just like entanglement, can be treated as a
physical resource,  and many coherence measures have been proposed \cite{coherence,rev1,rev2,rev3,cc,cos,yang,ma,negative1,robust}. So one may
ask whether methods to obtain lower bounds for entanglement measures
can also be used for coherence measures.

Experimentally quantifying entanglement and coherence are also extremely important. However, the quantum state tomography requires exponentially growing measurements with the number of qubits. To quantify GME in a large-qubit system such as 8-12 qubits, one needs a method to directly measure its GME, since it is difficult to perform quantum state tomography.

The purpose of this work is two-fold: On the one hand, we present
a method to obtain lower bounds
for GME measures of multipartite quantum states,
such as the convex-roof extended negativity of GME, the concurrence
of GME, the G-concurrence of GME, and the geometric measure of GME. On the
other hand we show that our method is also useful to obtain lower
bounds for coherence measures, such as the convex roof of the $l_{1}$-norm
of coherence, the geometric measure of coherence and the coherence of
formation. Our lower bounds are experimentally accessible without quantum state tomography, by determining the
fidelity between the chosen pure state and the target state.
Moreover, we present several examples
including real experimental states.

\section{Fidelity-based lower bounds for GME measures}
\begin{figure}
\begin{center}
\includegraphics[scale=0.7]{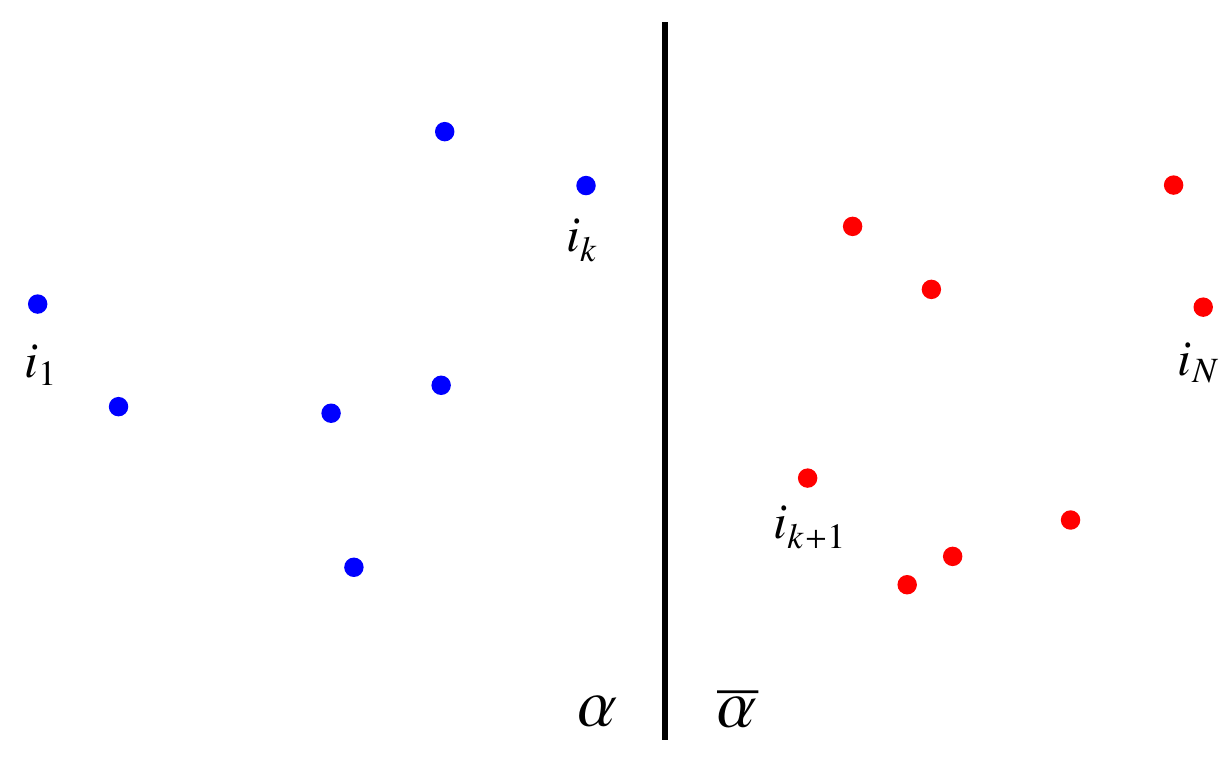}
\caption{Illustration for any possible bipartition. The index $\alpha$ denotes any possible bipartition $\alpha|\bar{\alpha}$  of $\{1,2,\cdots,N\}$. For instance, suppose that $i_1,i_2,\cdots,i_N$ is an arbitrary order of $1,2,\cdots,N$. The subset $\alpha$ contains $i_1,\cdots,i_k$ and $\bar{\alpha}=\{i_{k+1},\cdots,i_N\}$ with $1\leq k\leq N-1$, since $\alpha$ and $\bar{\alpha}$ are two nonempty subsets.} \label{fig1}
\end{center}
\end{figure}

For the quantification of multipartite entanglement, there exists a simple
way to generalize an arbitrary bipartite entanglement measure to an GME
measure. Considering a bipartite entanglement measure $E$, one can define
the GME measure as
\begin{equation}\label{}
   E_{\mathrm{GME}}(|\psi\rangle)=\min_{\alpha} E_{\alpha}(|\psi\rangle),
\end{equation}
for $N$-partite pure state $|\psi\rangle$, where $\alpha$ represents all possible
bipartitions $\alpha|\bar{\alpha}$ of $\{1,2,\cdots,N\}$, see Fig.~\ref{fig1}, and $E_{\alpha}(|\psi\rangle)$
is the bipartite entanglement measure $E$ under the bipartition $\alpha|\bar{\alpha}$.
The GME measure $E_{\mathrm{GME}}$ can be generalized to mixed states $\varrho$ via
a convex roof construction, i.e.,
\begin{equation}\label{crGME}
   E_{\mathrm{GME}}(\varrho)=\inf_{\{p_i,|\psi_i\rangle\}}\sum_i p_i E_{\mathrm{GME}}(|\psi_i\rangle),
\end{equation}
where the minimization runs over all possible decompositions $\varrho=\sum_i p_i|\psi_i\rangle\langle\psi_i|$. Clearly, evaluating this minimization is not
straightforward.

Before embarking on our main results, let us
recall some facts about an arbitrary $N$-partite pure entangled state $|\phi\rangle$,
which will be used in Theorems 1 to 4. As we mentioned above, $\alpha$
denotes any possible bipartition $\alpha|\bar{\alpha}$  of $\{1,2,\cdots,N\}$,  the
arbitrarily chosen $|\phi\rangle$ has its Schmidt decomposition
$|\phi\rangle=V_{\alpha}\otimes V_{\overline{\alpha}}
\sum_{i=1}^{m_\alpha}(s_i^{(\alpha)})^{\frac{1}{2}}|ii\rangle$ under the bipartition
$\alpha|\bar{\alpha}$, where $\{(s_i^{(\alpha)})^{\frac{1}{2}}\}$ are its Schmidt
coefficients in decreasing order and $m_\alpha$ is the total number of non-vanishing
Schmidt coefficients. We can define $s_{1}':=\max_{\alpha} \{s_1^{(\alpha)}\} $ and
$m':=\max_{\alpha} \{m_{\alpha}\}$, which can be easily calculated once $|\phi\rangle$
has been chosen.

For a bipartite pure state $|\psi\rangle$, the convex-roof extended negativity (CREN) is defined by the negativity
$\mathcal{N}(|\psi\rangle)=\||\psi\rangle\langle\psi|^{T_B}\|-1$ \cite{negativity,negativity21,CREN}.
For an $N$-partite pure state $|\psi\rangle$, the  CREN of GME can be defined as $\mathcal{N}_{\mathrm{GME}}(|\psi\rangle)=\min_{\alpha} \mathcal{N}_{\alpha}(|\psi\rangle)$,
and it can be generalized to mixed states by the convex roof. We can formulate:

\begin{theorem}
For any $N$-partite state $\varrho$, its convex-roof extended
negativity of GME satisfies
\begin{eqnarray}\label{NCRENkk}
\mathcal{N}_{\mathrm{GME}}(\varrho)\geq  S-1,
\end{eqnarray}
where $S=\max\{\langle\phi|\varrho|\phi\rangle/s_1',1\}$. Here, $|\phi\rangle$ is
an arbitrary pure state.
\end{theorem}

The proof is presented in the Appendix A. The interpretation of the lower bound in Theorem 1 is the following: If $\varrho$ has a high overlap
with some highly entangled state $|\phi\rangle$, this can be used to estimate the entanglement.
In experiments one only needs the fidelity with respect to $|\phi\rangle$.

The concurrence was introduced for two-qubit states in
Refs.~\cite{EOFde1,2qubit1,2qubit2}. In bipartite higher-dimensional
systems, the concurrence is defined by $C(|\psi\rangle)=\sqrt{2(1-\tr\varrho_A^2)}$
for pure states, and by the convex roof for mixed states
\cite{concurrence3,concurrence4,concurrence5}. Similar to the
convex-roof extended negativity of GME, one can get a lower bound
for the concurrence of GME in multipartite systems.

\begin{theorem}
For any $N$-partite state $\varrho$, a lower bound on the concurrence of GME
is given by
\begin{eqnarray}\label{concurrence2}
 C_{\mathrm{GME}}(\varrho)\geq \sqrt{\frac{2}{m'(m'-1)}}(S-1),
\end{eqnarray}
where $S=\max\{\langle\phi|\varrho|\phi\rangle/s_1',1\}$.
\end{theorem}

The proof is similar to Theorem 1, details are given in the Appendix A.

For an $m\otimes n$ ($m\leq n$) pure state $|\psi\rangle$, the G-concurrence
is defined by  $G(|\psi\rangle)=m(\det \varrho_A)^{1/m}$ \cite{GME1,Gour,Fan},
and by the convex roof for mixed states \cite{Uhlmann}. An interesting
feature of the G-concurrence is that it indicates entanglement of maximal
Schmidt rank, i.e., it vanishes for pure states where  $\varrho_A$ is not of
maximal rank. A lower bound for G-concurrence of GME can also be provided:

\begin{theorem}
For any $N$-partite state $\varrho$, its G-concurrence
of GME  satisfies
\begin{eqnarray}\label{gc2}
G_{\mathrm{GME}}(\varrho)\geq 1-m'+S,
\end{eqnarray}
where $S=\max\{\langle\phi|\varrho|\phi\rangle/s_1',1\}$.
\end{theorem}

The proof is similar to Theorem 1, details are given in the Appendix A.

\begin{figure}[htb]
\centering
\includegraphics[scale=0.6]{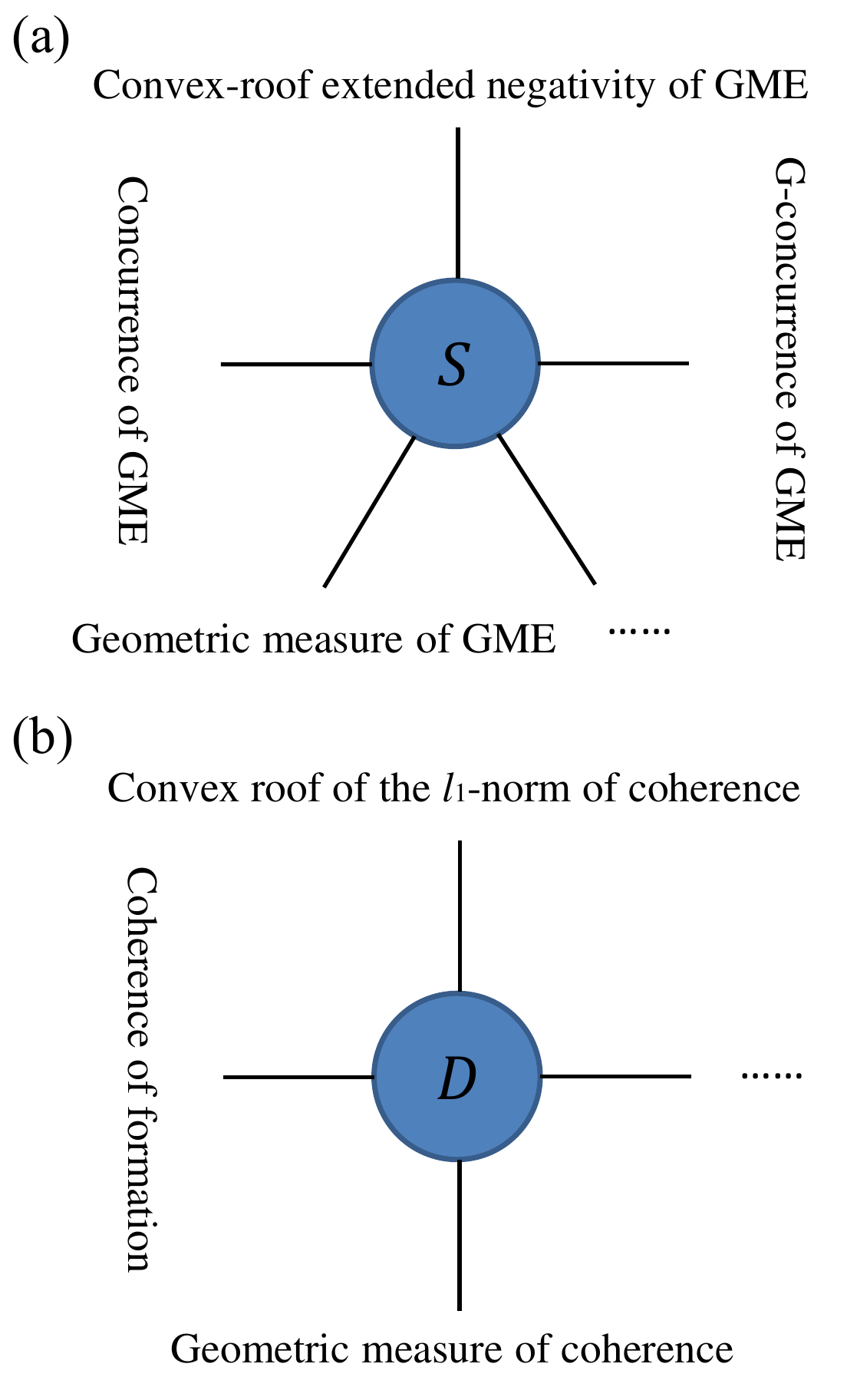}
\caption{Estimating GME and coherence measures via $S$ and $D$.
(a) All lower bounds of GME measures in Theorem 1-4 are related to $S$. In principle, one can obtain lower bounds for all convex-roof constructed GME measures Eq. (\ref{crGME}) from $S$.
(b) Convex-roof constructed coherence measures in Theorem 5-7 are related to $D$. Actually, $D$ is analogue to $S$, which is a simpler case in a single system.
}\label{fig2}
\end{figure}

For an arbitrary bipartite pure state $|\psi\rangle=U_A\otimes U_B \sum_i\sqrt{\mu_i}|ii\rangle$ with $\sqrt{\mu_i}$ being its Schmidt coefficients, the geometric measure of entanglement is defined by $\mathcal{G}(|\psi\rangle)=1-\max_i\{\mu_i\}$ \cite{GME1,GME2}. Similarly, the geometric measure of entanglement is extended to mixed states by the convex roof.
The last entanglement measure for which we provide the lower bound is the
geometric measure of GME.

\begin{theorem}
For an arbitrary $N$-partite state $\varrho$, its geometric
measure of GME $\mathcal{G}(\varrho)$ satisfies
\begin{eqnarray}\label{GMEgme2kk}
\mathcal{G}_{\mathrm{GME}}(\varrho)\geq 1-\gamma(S),
\end{eqnarray}
where $\gamma(S)=[\sqrt{S}+\sqrt{(m'-1)(m'-S)}]^2/m'^2$
with $S=\max\{\langle\phi|\varrho|\phi\rangle/s_1',1\}$.
\end{theorem}

The proof is similar to Theorem 1, details are given in the Appendix A.

From Theorem 1 to Theorem 4, all lower bounds provide a simple and convenient
way to evaluate GME for finite-dimensional multipartite states. Instead of finding
the minimal pure state decomposition, we only need to seek the maximal Schmidt
coefficient of the chosen state $|\phi\rangle$  among all possible bipartitions, and
obtain the fidelity between $|\phi\rangle$ and the target state $\varrho$.
It is remarkable that the choice of the state $|\phi\rangle$ is significant
for determining the bounds. Note that all the bounds are related to the entanglement witness \cite{Ewitness}
$W=s_1'\idol-|\phi\rangle\langle\phi|$, and this witness detects only states that
have a negative partial transpose (NPT) for any bipartition \cite{NPT}. On the
other hand, if this witness does not detect the states (and $S \leq 1$), then
there is a biseparable state compatible with the measured fidelity of
$|\phi\rangle$ and one cannot conclude that the state contained GME. In this
sense, our bounds are optimal. It would be interesting to study whether our
bounds also deliver the best possible value for $S>0$ in the sense of
Ref.~\cite{bounds17}.

\section{Fidelity-based lower bounds for coherence measures}
Before focusing on coherence measures, let us first compare the lower bounds of GME measures and coherence measures. As shown in Fig. \ref{fig2}, one can find that all the lower bounds of GME measures in Theorem 1-4 are related to $S$, which can be viewed as renormalized fidelity between the chosen pure state $|\phi\rangle$ and the target state $\varrho$. In the following, we will also find coherence measures in Theorem 5-7 are related to an analogue renormalized fidelity $D$.

For an $m$-dimensional pure state, a simple and reliable measure of coherence is the $l_{1}$-norm of coherence \cite{coherence}. It is defined as $C_{l_{1}}(\varrho)=\sum_{i\neq j}^m\big|\langle i|\varrho|j\rangle\big|$, i.e., the summation of all non-diagonal elements of density matrix. For mixed states, the convex roof can also be used in the $l_{1}$-norm measure \cite{cc}, just like entanglement measures. The convex roof of the $l_{1}$-norm is defined as $\widetilde{C}_{l_{1}}(\varrho)=\inf_{\{p_i,|\psi_i\rangle\}}\sum_i p_i C_{l_{1}}(|\psi_i\rangle)$ for all possible ensemble realizations of pure states $\varrho=\sum_i p_i |\psi_i\rangle\langle\psi_i|$. Our method can also be used for estimating
the coherence:

\begin{theorem}
For an arbitrary state $\varrho$ in an $m$-dimensional
system the convex roof of the $l_{1}$-norm of coherence satisfies
\begin{eqnarray}\label{l1norm}
\widetilde{C}_{l_{1}}(\varrho)\geq  D-1,
\end{eqnarray}
where $D=\max\{\langle\phi|\varrho|\phi\rangle/|d_{\mathrm{max}}|^2,1\}$,
and $|\phi\rangle=\sum_{i=1}^m d_{i}|i\rangle$ is an arbitrary $m$-dimensional
pure state with $|d_{\mathrm{max}}|=\max_i\{|d_{i}|\}$.
\end{theorem}

The proof is presented in the Appendix B.

The idea behind the geometric measure of coherence comes from the geometric
measure of entanglement. For a pure state $|\psi\rangle$, we have its geometric
measure of coherence $C_{g}(|\psi\rangle)=1-\max_{i}|\langle i|\psi\rangle|^2$.
For a general mixed state $\varrho$, the convex roof construction is used \cite{cos}.

\begin{theorem}
For an arbitrary state $\varrho$ in an $m$-dimensional
system the geometric measure of coherence satisfies
\begin{eqnarray}\label{GMEgme11}
C_g(\varrho)\geq 1-\gamma(D),
\end{eqnarray}
where $\gamma(D)=[\sqrt{D}+\sqrt{(m-1)(m-D)}]^2/m^2$, $D=\max\{\langle\phi|\varrho|\phi\rangle/|d_{\mathrm{max}}|^2,1\}$, and $|\phi\rangle=\sum_{i=1}^m d_{i}|i\rangle$ is an arbitrary $m$-dimensional pure state with $|d_{\mathrm{max}}|=\max_i\{|d_{i}|\}$.
\end{theorem}

The proof is similar to Theorem 5, details are given in
the Appendix B.

For a pure state $|\psi\rangle$, its coherence of formation is defined
as $C_{f}(|\psi\rangle)=S(\Delta(|\psi\rangle\langle\psi|))$, where $S$
is the von Neumann entropy and $\Delta(\varrho)=\sum_{i}|i\rangle\langle i|\varrho|i\rangle\langle i|$. For a general mixed state $\varrho$, the convex roof is used  \cite{yang,ma}. Moreover, it is proved that for any state $\varrho$ the coherence cost is given by the coherence of formation, i.e., $C_c(\varrho)=C_f(\varrho)$ \cite{yang}. Thus, the lower bound for coherence of formation is also the lower bound for coherence cost.

\begin{theorem}
For an arbitrary state $\varrho$ in an $m$-dimensional system, the
coherence of formation satisfies
\begin{eqnarray}\label{GMEgme11}
C_f(\varrho)\geq \mathcal{R}(D),
\end{eqnarray}
where
\begin{eqnarray}
\mathcal{R}(D)&=&\left\{
\begin{array}{ll}
H_2[\gamma(D)]+[1-\gamma(D)]\log_2(m-1), \\
\mathrm{when} \ \ D\in[1,\frac{4(m-1)}{m}]; \\
(D-m)\frac{\log_2(m-1)}{m-2}+\log_2 m, \\
\mathrm{when} \ \ D\in[\frac{4(m-1)}{m},m].
\end{array}%
\right.
\end{eqnarray}
$H_2(x)=-x\log_2 x-(1-x)\log_2(1-x)$, $\gamma(D)=[\sqrt{D}+\sqrt{(m-1)(m-D)}]^2/m^2$, $D=\max\{\langle\phi|\varrho|\phi\rangle/|d_{\mathrm{max}}|^2,1\}$, and $|\phi\rangle=\sum_{i=1}^m d_{i}|i\rangle$ is an arbitrary $m$-dimensional pure state with $|d_{\mathrm{max}}|=\max_i\{|d_{i}|\}$.
\end{theorem}

The proof is similar to Theorem 5, details are given in
the Appendix B.

Note that it is not surprising that the form of the $l_{1}$-norm bound is similar to the CREN
bound. Ref.~\cite{negative1} suggested that for pure states the negativity of $\sum_j\sqrt{\lambda_j}|jj\rangle$ is equal to the $l_{1}$-norm of coherence for $\sum_j\sqrt{\lambda_j}|j\rangle$, and the two measures share the same convex roof. The lower bounds for geometric measure of coherence is also similar to the lower bounds for geometric measure of entanglement. Ref.~\cite{yang} pointed out that for a general state $\varrho=\sum_{ij}\varrho_{ij}|i\rangle\langle j|$, the coherence of formation is equal to the entanglement of formation of the corresponding maximally correlated state $\varrho_{mc}=\sum_{ij}\varrho_{ij}|ii\rangle\langle jj|$. So the observable lower bounds are naturally extended from entanglement to coherence.

Similar to the entanglement witness, the concept of coherence witness has been proposed in Ref. \cite{robust}.
From Theorem 5 to Theorem 7, all the observable lower
bounds of coherence measures are related to the same coherence witness, i.e. $W=|d_{\mathrm{max}}|^2\idol-|\phi\rangle\langle\phi|$, where $|\phi\rangle=\sum_{i=1}^m d_{i}|i\rangle$ is an arbitrary
$m$-dimensional pure state with $|d_{\mathrm{max}}|=\max_i\{|d_{i}|\}$ and $\{|i\rangle\}$
is the reference basis. For all the incoherent states $\delta=\sum_i p_i |i\rangle\langle i|$,
we have $\tr(W\delta)=|d_{\mathrm{max}}|^2-\sum_i p_i |d_i|^2\geq0$. If there exists a state
$\varrho$ such that $\tr(W\varrho)<0$, the state $\varrho$ must be a coherent state.
Finally, concerning other coherence measures than the ones from above, a proposal has recently
been made to estimate the relative entropy of coherence \cite{xiaodong}.

\begin{figure}
\begin{center}
\includegraphics[scale=0.85]{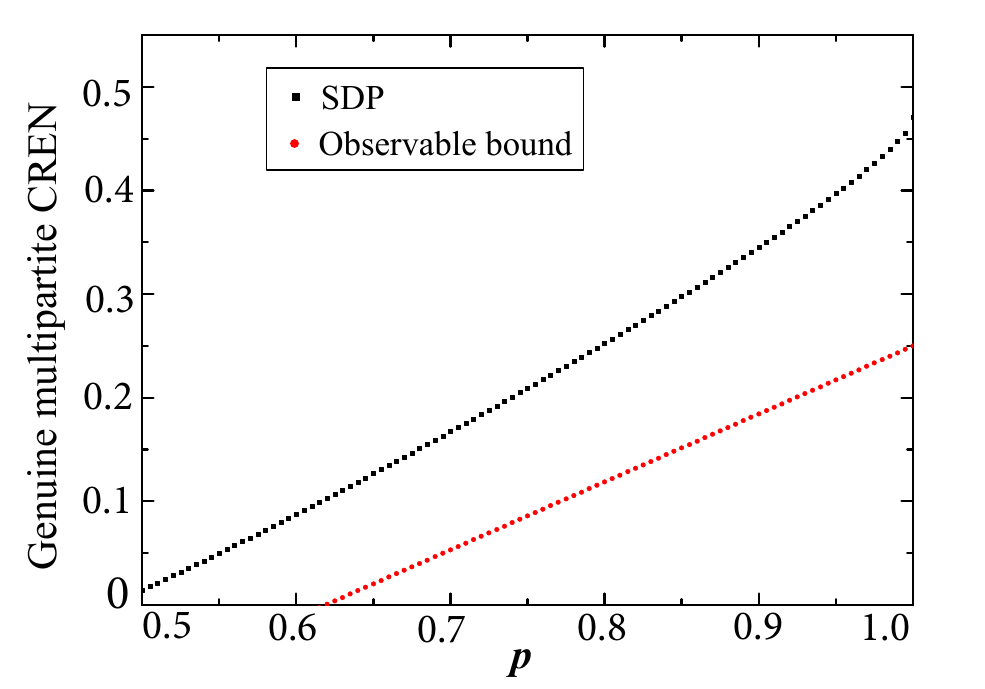}
\caption{{\bf Compare SDP results with the GME observable lower bound.}
The results of SDP and the GME observable lower bound are shown here. The black square dots denote the renormalized GMN of $\varrho_{W}$ by using SDP, where $\varrho_{W}=p|W\rangle\langle W|+(1-p)\idol/8$. The red dots denote the GME observable lower bound of CREN for $\varrho_{W}$ based on Eq. (\ref{NCRENkk}), where we have used $|\phi\rangle=|W\rangle$ with $s_1'=2/3$. The mixed coefficient $p$ is plotted on the horizontal axis while the entanglement measures are plotted on the vertical axis.} \label{fig3}
\end{center}
\end{figure}

\section{Examples}
We provide two examples to contrast our results with the results from genuine multiparticle negativity (GMN) in Ref.~\cite{negative2}.

The GMN measurement was built on the analysis of bipartite decompositions of the multipartite state $\varrho$. If we can find an entanglement witness $W$ which is decomposable for all possible bipartite decompositions $\alpha|\overline{\alpha}$ and it can detect the state $\tr(\varrho W)<0$ , the state $\varrho$ is GME state. Here $W=P_{\alpha}+Q_{\alpha}^{T_{\alpha}}$ with positive operators $P_{\alpha}$ and $Q_{\alpha}$. The GMN was defined as $\widetilde{N}_{g}(\varrho)=-\min \tr (\varrho W)$, where $0\leq P_{\alpha}\leq \idol$ and $0\leq Q_{\alpha}\leq \idol$. Base on the GMN, Ref. \cite{negative2} proposed the renormalized GMN, where the operator $P$ is not bounded by $\idol$ anymore. Interestingly, the renormalized GMN is equal to a mixed-state convex roof of bipartite negativity:
\begin{eqnarray}\label{GMN}
N_g(\varrho)=\inf_{p_\alpha,\varrho_{\alpha}}\sum_\alpha p_\alpha N_{\alpha}(\varrho_{\alpha}),
\end{eqnarray}
where the summation runs over all possible decompositions $\alpha|\overline{\alpha}$ of the system and the minimization is performed over all mixed state decompositions of the state $\varrho=\sum_\alpha p_\alpha \varrho_{\alpha}$. Different from the pure convex roof, such as the CREN, this definition is built on the mixed state decomposition. The mixed state decompositions also include the pure state decompositions. Therefore the bound described by the mixed convex roof is a lower bound of the pure-state convex roof.

\textit{Example 1.} Considering a $n$-qubit GHZ-diagonal state $\varrho$, its renormalized GMN $N_g(\varrho)$ satisfies $N_g(\varrho)= \max_{i}\{2F_{i}-1,0\}$ (we use a different prefactor compared with Ref. \cite{negative2}), where $F_{i}=\langle\psi_i|\varrho|\psi_i\rangle$ denotes the fidelity with the GHZ-basis state $|\psi_i\rangle$.

Our bound on the convex-roof extended negativity of GME can also give the same result. We adopt the GHZ-basis state with the maximal fidelity as the observable state $|\phi\rangle$ then $s_1$ can only the value $1/2$. Therefore, Eq. (\ref{NCRENkk}) reads as $\mathcal{N}(\varrho)\geq\max\{2\langle\psi_i|\varrho|\psi_i\rangle-1,0\}$. This lower bound is just equal to the analytical result in Ref. \cite{negative2}.

\begin{figure*}[hbtp]
\includegraphics[width=\textwidth]{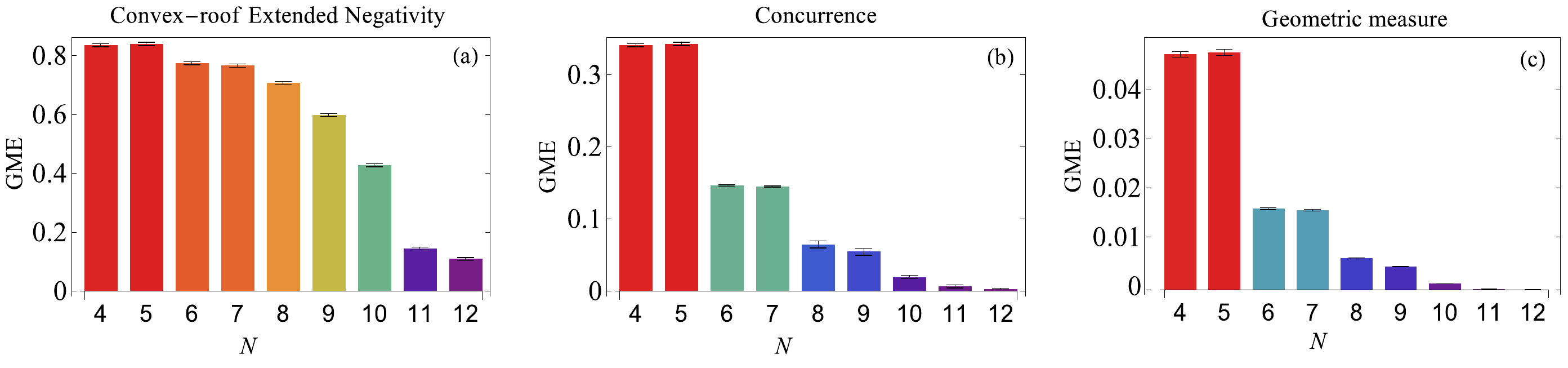}
\caption{Lower bounds of GME measures for linear cluster states.
(a) The lower bound of convex-roof extended negativity of GME versus the qubit number from 4 to 12.
(b) The lower bound of concurrence of GME versus the qubit number from 4 to 12.
(c) The lower bound of geometric measure of GME versus the qubit number from 4 to 12.
}\label{fig4}
\end{figure*}

\textit{Example 2.} Consider a $3$-qubit $W$ state with white noise $\varrho_{W}=p|W\rangle\langle W|+(1-p)\idol/8$, where $|W\rangle=1/\sqrt{3}(|001\rangle+|010\rangle+|100\rangle)$. We measure its entanglement respectively with the renormalized GMN and convex-roof extended negativity of GME. A semidefinite program (SDP) proposed in Ref. \cite{negative2} is used here to calculate the renormalized GMN. The result is showed in Fig. \ref{fig3}. The GME lower bound is less than the SDP result for each $p$. The advantage of the GME lower bound is that we only need to measure the expectation value $\langle\phi|\varrho|\phi\rangle$ rather than detect $\varrho$ itself.

The genuine multiparticle negativity in Ref.~\cite{negative2} needs quantum state tomography. In the following, we provide two examples using real experimental data in which quantum state tomography is usually impractical.

\textit{Example 3.} We adopt three experimental results of multipartite systems \cite{Nph1,Nph2,Nph3} to calculate the GME lower bound. All the three experiments are experimental
demonstrations of multipartite quantum entanglement. The produced states in these experiments achieve $6$-$10$ photons and it is very hard to perform quantum tomography. In these experiments, coincidence counts measured in the $|H\rangle/|V\rangle$ basis and the fidelity with $N$-GHZ state is used to describe the produced states. In these cases, It is impossible to get exact results of most entanglement measures. Our method provides a reliable way to solve the question. From Table \ref{table1}, one can see that the genuine entanglement exists in the experimentally realized states of Refs. \cite{Nph1,Nph2,Nph3}, quantified by most entanglement measures except the G-concurrence of GME. We can also estimate the coherence of the produced states with the lower bound.

\begin{table}[h]
\caption{Results of entanglement and coherence measures on $N$-photon entanglement states. The high corner marks $a$, $b$, $c$ denote different data sources of Refs. \cite{Nph1,Nph2,Nph3}, respectively. The fidelity is the result of overlap $\langle\phi|\varrho|\phi\rangle$ with $|\phi\rangle=1/\sqrt{2}(|H\rangle^{\otimes N}+|V\rangle^{\otimes N})$}. \label{table1}
\begin{tabular}{c|c c c c c}
\hline
\hline
   $\varrho(N)$  & $6^a$ & $8^a$ & $8^b$ & $8^c$ & $10^a$ \\
\hline
Fidelity  & $0.710(16)$  & $0.644(22)$ &$0.610(26)$ & $0.59(2)$ &$0.573(23)$ \\
\hline
$\mathcal{N}_{\mathrm{GME}}(\varrho)$  & $0.420$ & $0.288$ & $0.220$ & $0.18$ & $0.146$ \\

$C_{\mathrm{GME}}(\varrho)$ &$0.0794$& $0.0263$ & $0.020$1 & $0.016$ & $0.0066$ \\

$G_{\mathrm{GME}}(\varrho)$ & $0$ & $0$ & $0$ & $0$ & $0$ \\

$\mathcal{G}_{\mathrm{GME}}(\varrho)$  & $0.00540$ & $0.00122$ &  $0.00073$ &  $0.00049$ & $0.00016$ \\
\hline
$\widetilde{C}_{l_{1}}(\varrho)$ & $0.420$ & $0.288$ & $0.220$ & $0.18$ & $0.146$\\

$C_g(\varrho)$ &  5.85e-4 & 7.14e-5 & 4.29e-5 & 2.9e-5 & 4.86e-6 \\

$C_f(\varrho)$ & 0.0106   & 0.00165 & 0.00102 & 7.1e-4 & 1.41e-4 \\
\hline
\hline
\end{tabular}
\end{table}

\textit{Example 4.} We consider the experiment of a genuine 12-qubit entanglement in a superconducting processor in Ref. \cite{exp12}. Linear cluster states have been prepared from 4 to 12 qubits by implementing a set of controlled-phase gates on the superconducting quantum processor. Since the quantum state tomography requires exponentially growing measurements with the number of qubits, the lower bounds of the state fidelities are measured with probability distributions, which are 0.9176(28), 0.9196(28), 0.8870(27), 0.8827(27), 0.8536(27), 0.7988(27), 0.7136(26), 0.5720(25), and 0.5544(25), from 4 to 12 qubits \cite{exp12}. It is worth noting that the right hand sides of Eqs. (\ref{NCRENkk}), (\ref{concurrence2}), (\ref{gc2}) and  (\ref{GMEgme2kk}) are monotonically increasing functions of $S$. Thus, if we use a lower bound of fidelity in $S$, one can still obtain a valid lower bound of GME measure. In Fig. \ref{fig4}, the lower bounds of GME measures including CREN of GME, concurrence of GME and geometric measure of GME have been shown for the experimentally realized linear cluster states from 4 to 12 qubits.

\section{Discussions and conclusions}
The observable lower bounds for the entanglement of bipartite states
have been discussed in Ref.~\cite{observe}, and we have extended it to lower bounds for
GME and coherence. Unlike the constant $m$ and $s_{1}$ for bipartite states,
$m_{\alpha}$ and $s_{1}^{(\alpha)}$ often change with different bipartite decompositions.
For calculating GME lower bounds, the key point is to find the maximal $m_\alpha$ and
$s_{1}^{(\alpha)}$ among all bipartite decompositions $\alpha|\overline{\alpha}$. Although
the fidelity and Schmidt coefficients are related to the entanglement witness and the geometric measure
of entanglement, our lower bounds of GME are substantially different from previous entanglement lower bounds.

The form of the lower bounds for coherence is very similar to the bounds for entanglement.
This indicates a strong relation between entanglement and coherence. In certain cases, we
can measure coherence just as measure entanglement. We find this relation in the
$l_{1}$-norm of coherence, the geometric measure of coherence, and the coherence of formation.
It is an open question whether other coherence measures can apply for such an observable lower
bound.

In conclusion, we proposed a method to obtain lower bounds for some genuine entanglement
and coherence measures defined by the convex roof construction. The entanglement measures
include the convex-roof extended negativity, the concurrence, the G-concurrence and
the geometric measure of entanglement, while coherence measures include the convex roof
of $l_{1}$-norm of coherence, the geometric measure of coherence, and the coherence of
formation. The lower bounds estimate these measures for arbitrary finite-dimensional
multipartite states. Moreover, these lower bounds can be easily obtained from the fidelity between
a chosen pure state $|\phi\rangle$ and the target state.
For future work it would be very interesting to extend our results to other figures of merit in
quantum information, such as distillability rates, or the usefulness of a state for tasks
like teleportation or metrology.

\section*{ACKNOWLEDGMENTS}
C.Z. gratefully acknowledges the support of the K.~C.~Wong Education Foundation and the DAAD.
This work is partially supported by the National Natural Science Foundation of China
(Grants No. 11674238 and No. 11734015). O.G. gratefully acknowledges the support of the DFG and the ERC
(Consolidator Grant 683107/TempoQ).

\section*{APPENDIX A: PROOFS OF THEOREM 1-4}
\textit{Theorem 1.} For any $N$-partite state $\varrho$, its convex-roof extended
negativity of GME satisfies
\begin{eqnarray}\label{NCREN}
\mathcal{N}_{\mathrm{GME}}(\varrho)\geq  S-1,
\end{eqnarray}
where $S=\max\{\langle\phi|\varrho|\phi\rangle/s_1',1\}$. Here, $|\phi\rangle$ is
an arbitrary pure state.

\textit{Proof.}  Assume that $\varrho=\sum_j p_j|\psi_j\rangle\langle\psi_j|$ is
the optimal decomposition for $\varrho$ to achieve the infimum of $\mathcal{N}_{\mathrm{GME}}(\varrho)=\inf_{\{p_i,|\psi_i\rangle\}}\sum_i p_i \mathcal{N}_{\mathrm{GME}}(|\psi_i\rangle)$,
and then $\mathcal{N}_{\mathrm{GME}}(\varrho)=\sum_j p_j \mathcal{N}_{\mathrm{GME}}(|\psi_j\rangle)$.
For all possible bipartitions $\alpha|\bar{\alpha}$, we can calculate the CREN of each
$|\psi_j\rangle$: $\mathcal{N}_{\alpha}(|\psi_j\rangle)=\||\psi_j\rangle\langle\psi_j|^{T_\alpha}\|-1=\big(\sum_{i=1}^{m_\alpha}(\mu_i^{(j, \alpha)})^{\frac{1}{2}}\big)^2-1$,
where $(\mu_i^{(j, \alpha)})^{\frac{1}{2}}$ are Schmidt coefficients of $|\psi_j\rangle$
in decreasing order under the bipartition $\alpha|\overline{\alpha}$, and $\|\cdot\|$ is
the trace norm. Then we have
\begin{eqnarray}\label{NCREN2}
&&\mathcal{N}_{\mathrm{GME}}(|\psi_j\rangle)=\min_{\alpha}\mathcal{N}_{\alpha}(|\psi_j\rangle)
\nonumber\\
&& \geq\min_{\alpha}\frac{\langle\phi|\psi_j\rangle\langle\psi_j|\phi\rangle}{s_1^{(\alpha)}}-1\geq\frac{\langle\phi|\psi_j\rangle\langle\psi_j|\phi\rangle}{s_1'}-1,
\end{eqnarray}
where we have used the following inequality
\begin{eqnarray}
\Bigg(\sum_{i=1}^{m_\alpha}\sqrt{\mu_i^{(j, \alpha)}}\Bigg)^2\geq\frac{\langle\phi|\psi_{j}\rangle\langle\psi_{j}|\phi\rangle}{s_{1}^{(\alpha)}},
\end{eqnarray}
which can be proved according to Refs.~\cite{nonlinear,observe}. Then we can find the
GME lower bound
\begin{eqnarray}\label{NCREN3}
&&\mathcal{N}_{\mathrm{GME}}(\varrho)= \sum_j p_j\mathcal{N}_{\mathrm{GME}}(|\psi_{j}\rangle)\geq S-1.
\end{eqnarray}
Thus, Theorem 1 has been proved.

\textit{Theorem 2.} For any $N$-partite state $\varrho$, its GME lower bound of concurrence  satisfies
\begin{eqnarray}
 C_{\mathrm{GME}}(\varrho)\geq \sqrt{\frac{2}{m'(m'-1)}}(S-1),
\end{eqnarray}
where $S=\max\{\langle\phi|\varrho|\phi\rangle/s_1',1\}$.

\textit{Proof.} Similarly, suppose that $\varrho=\sum_j p_j|\psi_j\rangle\langle\psi_j|$ is the optimal decomposition for $\varrho$ to achieve the infimum of $C_{\mathrm{GME}}(\varrho)=\inf_{\{p_i,|\psi_i\rangle\}}\sum_i p_i C_{\mathrm{GME}}(|\psi_i\rangle)$, and then $C_{\mathrm{GME}}(\varrho)=\sum_j p_j C_{\mathrm{GME}}(|\psi_j\rangle)$. For the bipartite decomposition $\alpha|\overline{\alpha}$, we can calculate the concurrence $C_\alpha(|\psi_j\rangle)=\sqrt{2[1-\sum_{i=1}^{m_\alpha}(\mu_i^{(j, \alpha)})^2]}$ with $(\mu_i^{(j, \alpha)})^{\frac{1}{2}}$ being Schmidt coefficients of $|\psi_j\rangle$ in decreasing order under the bipartition $\alpha|\overline{\alpha}$. Thus,
\begin{eqnarray}
C_{\mathrm{GME}}(|\psi_j\rangle)&=& \min_\alpha C_\alpha(|\psi_j\rangle)\nonumber\\
&\geq&\min_\alpha \sqrt{\frac{2}{m_\alpha(m_\alpha-1)}}\big(\frac{\langle\phi|\psi_j\rangle\langle\psi_j|\phi\rangle} {s_1^{(\alpha)}}-1\big)\nonumber\\
&\geq&\sqrt{\frac{2}{m'(m'-1)}}\big(\frac{\langle\phi|\psi_j\rangle\langle\psi_j|\phi\rangle} {s_1'}-1\big),
\end{eqnarray}
where the first inequality is proved in Ref. \cite{observe}. Therefore,
\begin{equation}
C_{\mathrm{GME}}(\varrho)=\sum_j p_j C_{\mathrm{GME}}(|\psi_j\rangle)\geq\sqrt{\frac{2}{m'(m'-1)}}(S-1).
\end{equation}
Thus, Theorem 2 has been proved.

\textit{Theorem 3.} For any $N$-partite state $\varrho$, its G-concurrence of GME  satisfies
\begin{eqnarray}
G_{\mathrm{GME}}(\varrho)\geq 1-m'+S,
\end{eqnarray}
where $S=\max\{\langle\phi|\varrho|\phi\rangle/s_1',1\}$.

\textit{Proof.} Similarly, suppose that $\varrho=\sum_j p_j|\psi_j\rangle\langle\psi_j|$ is the optimal decomposition for $\varrho$ to achieve the infimum of $G_{\mathrm{GME}}(\varrho)=\inf_{\{p_i,|\psi_i\rangle\}}\sum_i p_i G_{\mathrm{GME}}(|\psi_i\rangle)$, and thus $G_{\mathrm{GME}}(\varrho)=\sum_j p_j G_{\mathrm{GME}}(|\psi_j\rangle)$. For the bipartite decomposition $\alpha|\overline{\alpha}$, one obtains $G_{\alpha}(|\psi_j\rangle)= m_\alpha\big(\prod_{i=1}^{m_\alpha} \mu_i^{(j,\alpha)}\big)^{\frac{1}{m_\alpha}}$. Similar to the proofs of Theorem $1$, $2$, we can find the inequality following the proof of bipartite systems \cite{observe}:
\begin{eqnarray}
G_{\mathrm{GME}}(|\psi_j\rangle)&=& \min_{\alpha} G_{\alpha}(|\psi_j\rangle)\nonumber\\
&\geq&\min_{\alpha}(1-m_\alpha+\frac{\langle\phi|\psi_j\rangle\langle\psi_j|\phi\rangle} {s_1^{(\alpha)}})\nonumber\\
&\geq&1-m'+\frac{\langle\phi|\psi_j\rangle\langle\psi_j|\phi\rangle} {s_1'}.
\end{eqnarray}
Therefore, we obtain the lower bound,
\begin{eqnarray}
G_{\mathrm{GME}}(\varrho)=\sum_j p_j G_{\mathrm{GME}}(|\psi_j\rangle)\geq1-m'+S,
\end{eqnarray}
where $m'=\max_{\alpha}\{m_{\alpha}\}$ and $s_1'=\max_{\alpha}\{s_1^{(\alpha)}\}$.

Before the proof of Theorem 4, let us prove a lemma first.

\textit{Lemma 1.} The function $\gamma(S)=[\sqrt{S/m'}+\sqrt{(m'-1)(1-S/m')}]^2/m'$ with $S=\max\{\langle\phi|\rho|\phi\rangle/s_1',1\}$ is a monotone increasing function of $s_1'$ and $m'$.

\textit{Proof.} Firstly, we prove that $\gamma$ is a monotone increasing function of $s_1'$. For simplicity, let $S=\max\{\langle\phi|\varrho|\phi\rangle/s_1',1\}$. We can find that
\begin{widetext}
\begin{eqnarray}\label{app11}
\frac{\partial\gamma}{\partial S}=\big(\sqrt{(1-\frac{S}{m'})(m'-1)}+\sqrt{\frac{S}{m'}}\big)\big(\frac{1-m'}{\sqrt{(1-\frac{S}{m'})(m'-1)}}+\sqrt{\frac{m'}{S}}\big)/m'^2.
\end{eqnarray}
\end{widetext}
For determining the monotone of $\gamma(S)$, we only need to discuss $(1-m')/\sqrt{(1-S/m')(m'-1)}+\sqrt{m'/S}$. After simplifying the function, the condition of $\partial\gamma/\partial S<0 $ is $S> 1$. Therefore $\gamma$ is a monotone decreasing function of $S$, i.e. a monotone increasing function of $s_1'$.

Then we prove that $\gamma$ is a monotone increasing function of $m'$. Similarly, we calculate the partial derivative of $\gamma$ with respect to $m'$
\begin{widetext}
\begin{eqnarray}\label{app12}
\frac{\partial\gamma}{\partial m'}=(\sqrt{(1-\frac{S}{m'})(m'-1)}+\sqrt{\frac{S}{m'}})(\frac{\frac{S}{m'}(m'-2)+1}{\sqrt{(1-\frac{S}{m'})(m'-1)}}-2\sqrt{\frac{S}{m'}})/m'^2.
\end{eqnarray}
\end{widetext}
As the above proof, we only consider $\frac{(S/m')(m'-2)+1}{\sqrt{(1-\frac{S}{m'})(m'-1)}}-2\sqrt{\frac{S}{m'}}$. After simplifying, the condition of $\partial\gamma/\partial m'>0 $ is $(S-1)^2>0$. Therefore $\gamma$ is a monotone increasing function of $m'$.

\textit{Theorem 4.} For an arbitrary $N$-partite state $\varrho$, its geometric measure of GME $\mathcal{G}(\varrho)$ satisfies
\begin{eqnarray}\label{GMEgme2}
\mathcal{G}_{\mathrm{GME}}(\varrho)\geq 1-\gamma(S),
\end{eqnarray}
where $\gamma(S)=[\sqrt{S}+\sqrt{(m'-1)(m'-S)}]^2/m'^2$ with $S=\max\{\langle\phi|\varrho|\phi\rangle/s_1',1\}$.

\textit{Proof.} Similarly, suppose that $\varrho=\sum_j p_j|\psi_j\rangle\langle\psi_j|$ is the optimal decomposition for $\varrho$ to achieve the infimum of $\mathcal{G}_{\mathrm{GME}}(\varrho)=\inf_{\{p_i,|\psi_i\rangle\}}\sum_i p_i \mathcal{G}_{\mathrm{GME}}(|\psi_i\rangle)$. Thus, $\mathcal{G}_{\mathrm{GME}}(\varrho)=\sum_j p_j \mathcal{G}_{\mathrm{GME}}(|\psi_j\rangle)$. For the bipartition $\alpha|\overline{\alpha}$, one obtains $\mathcal{G}_\alpha(|\psi_j\rangle)=1-\mu_{\mathrm{max}}^{(j, \alpha)}$ where $\mu_{\mathrm{max}}^{(j, \alpha)}=\max_i\{\mu_i^{(j, \alpha)}\}$. Therefore,
\begin{eqnarray}\label{GMEgme3}
&&\mathcal{G}_{\mathrm{GME}}(|\psi_j\rangle)=\min_{\alpha}\mathcal{G}_{\alpha}(|\psi_j\rangle)\nonumber\\
&&\ \ \ \geq 1-\min_{\alpha}\frac{[\sqrt{S^{(j)}_\alpha}+\sqrt{(m_\alpha-1)(m_\alpha-S^{(j)}_\alpha)}]^2}{m_\alpha^2}\nonumber\\
&&\ \ \ \geq 1-\frac{[\sqrt{S^{(j)}}+\sqrt{(m'-1)(m'-S^{(j)})}]^2}{m'^2},
\end{eqnarray}
where the first inequality is proved in Ref.~\cite{observe}, $S^{(j)}_\alpha=\max\{\langle\phi|\psi_j\rangle\langle\psi_j|\phi\rangle/s_1^{(\alpha)},1\}$ and $S^{(j)}=\max\{\langle\phi|\psi_j\rangle\langle\psi_j|\phi\rangle/s_1',1\}$. In Lemma 1, we have proved that $\gamma(S)$ is a monotone increasing function of $m'$ and $s_1'$. Therefore, the last inequality of Eq. ({\ref{GMEgme3}}) holds. Ref. \cite{observe} proved that $1-\gamma(S)$ is a convex function, so $1-\sum_j p_j\gamma(S^{(j)})\geq1-\gamma(\sum_{j}p_j S^{(j)})$. Thus,
\begin{eqnarray}\label{GMEgme4}
\mathcal{G}_{\mathrm{GME}}(\varrho)&=&\sum_j p_j \mathcal{G}_{\mathrm{GME}}(|\psi_j\rangle)\geq1-\sum_{j}p_{j}\gamma(S^{(j)})\nonumber\\
&\geq& 1-\gamma(\sum_{j}p_{j}S^{(j)})=1-\gamma(S).
\end{eqnarray}

\section*{APPENDIX B: PROOFS OF THEOREM 5-7}
\textit{Theorem 5.} For an arbitrary state $\varrho$ in an $m$-dimensional
system the convex roof of the $l_{1}$-norm of coherence satisfies
\begin{eqnarray}\label{l1norm}
\widetilde{C}_{l_{1}}(\varrho)\geq  D-1,
\end{eqnarray}
where $D=\max\{\langle\phi|\varrho|\phi\rangle/|d_{\mathrm{max}}|^2,1\}$,
and $|\phi\rangle=\sum_{i=1}^m d_{i}|i\rangle$ is an arbitrary $m$-dimensional
pure state with $|d_{\mathrm{max}}|=\max_i\{|d_{i}|\}$.

\textit{Proof.}  Similarly, one can assume that $\varrho=\sum_{j}p_{j}|\psi_{j}\rangle\langle\psi_{j}|$ is the optimal decomposition for $\varrho$ to achieve the infimum of $\widetilde{C}_{l_{1}}(\varrho)=\inf_{\{p_i,|\psi_i\rangle\}}\sum_i p_i C_{l_{1}}(|\psi_i\rangle)$, where $|\psi_{j}\rangle=\sum_{k}a_{k}^{(j)}|k\rangle$. Therefore, $\widetilde{C}_{l_{1}}(\varrho)=\sum_{j}p_{j}C_{l_{1}}(|\psi_{j}\rangle)$. Each $C_{l_{1}}(|\psi_{j}\rangle)$ can be calculated as follows,
\begin{eqnarray}\label{l1norm1}
&&C_{l_{1}}(|\psi_{j}\rangle)=\sum_{l\neq k}|a_{l}^{(j)}||a_{k}^{(j)}|=(\sum_{k}|a_{k}^{(j)}|)^2-1\nonumber\\
&&\ \ \ \geq\frac{(\sum_{k}|d_{k}||a_{k}^{(j)}|)^2}{|d_{\mathrm{max}}|^2}-1\geq\frac{\langle\phi|\psi_{j}\rangle\langle\psi_{j}|\phi\rangle}{|d_{\mathrm{max}}|^2}-1.
\end{eqnarray}
Therefore, one has
\begin{eqnarray}\label{l1norm2}
\widetilde{C}_{l_{1}}(\varrho)&\geq& \frac{\sum_{j}p_{j}\langle\phi|\psi_{j}\rangle\langle\psi_{j}|\phi\rangle}{|d_{\mathrm{max}}|^2}-1=D-1.
\end{eqnarray}
Therefore, we have proved Theorem 5.

\textit{Theorem 6.} For an arbitrary state $\varrho$ in an $m$-dimensional system, its geometric measure of coherence satisfies
\begin{eqnarray}\label{GMEgme11}
C_g(\varrho)\geq 1-\gamma(D),
\end{eqnarray}
where $\gamma(D)=[\sqrt{D}+\sqrt{(m-1)(m-D)}]^2/m^2$, $D=\max\{\langle\phi|\varrho|\phi\rangle/|d_{\mathrm{max}}|^2,1\}$, and $|\phi\rangle=\sum_{i=1}^m d_{i}|i\rangle$ is an arbitrary $m$-dimensional pure state with $|d_{\mathrm{max}}|=\max_i\{|d_{i}|\}$.

\textit{Proof.} Similarly, suppose that $\varrho=\sum_{j}p_{j}|\psi_{j}\rangle\langle\psi_{j}|$ is the optimal decomposition for $\varrho$ to achieve the infimum of $\inf_{\{p_i,|\psi_i\rangle\}}\sum_i p_i C_g(|\psi_i\rangle)$. Thus, $C_g(\varrho)=\sum_j p_j C_g(|\psi_j\rangle)$. Similar to the proof of lower bound for geometric measure of entanglement in Ref. \cite{observe}, each $C_g(|\psi_j\rangle)$  can be calculated as $C_g(|\psi_j\rangle)\geq  1- \gamma(D^{(j)})$,
where $D^{(j)}=\max\{\langle\phi|\psi_j\rangle\langle \psi_j|\phi\rangle/|d_{\mathrm{max}}|^2,1\}$.
As  proved in Ref. \cite{observe}, $1-\gamma(D)$ is a convex function. Thus,
\begin{eqnarray}
C_g(\varrho)&=&\sum_j p_j C_g(|\psi_j\rangle)\geq1-\sum_j p_j \gamma(D^{(j)})\nonumber\\
&\geq&1-\gamma(\sum_j p_j D^{(j)})=1-\gamma(D).
\end{eqnarray}

\textit{Theorem 7.} For an arbitrary state $\varrho$ in an $m$-dimensional system, its coherence of formation satisfies
\begin{eqnarray}\label{GMEgme11}
C_f(\varrho)\geq \mathcal{R}(D),
\end{eqnarray}
where
\begin{eqnarray}
\mathcal{R}(D)&=&\left\{
\begin{array}{ll}
H_2[\gamma(D)]+[1-\gamma(D)]\log_2(m-1), \\
\mathrm{when} \ \ D\in[1,\frac{4(m-1)}{m}]; \\
(D-m)\frac{\log_2(m-1)}{m-2}+\log_2 m, \\
\mathrm{when} \ \ D\in[\frac{4(m-1)}{m},m].
\end{array}%
\right.
\end{eqnarray}
$H_2(x)=-x\log_2 x-(1-x)\log_2(1-x)$, $\gamma(D)=[\sqrt{D}+\sqrt{(m-1)(m-D)}]^2/m^2$, $D=\max\{\langle\phi|\varrho|\phi\rangle/|d_{\mathrm{max}}|^2,1\}$, and $|\phi\rangle=\sum_{i=1}^m d_{i}|i\rangle$ is an arbitrary $m$-dimensional pure state with $|d_{\mathrm{max}}|=\max_i\{|d_{i}|\}$.

\textit{Proof.} Similarly, suppose that $\varrho=\sum_{j}p_{j}|\psi_{j}\rangle\langle\psi_{j}|$ is the optimal decomposition for $\varrho$ to achieve the infimum of
$\inf_{\{p_i,|\psi_i\rangle\}}\sum_i p_i C_g(|\psi_i\rangle)$. Thus, $C_f(\varrho)=\sum_j p_j C_f(|\psi_j\rangle)$. Similar to the proof of the lower bound for  entanglement of formation in Ref. \cite{observe}, each $C_f(|\psi_j\rangle)$  can be calculated as $C_f(|\psi_j\rangle)\geq \mathcal{R}(D^{(j)})$,
where $D^{(j)}=\max\{\langle\phi|\psi_j\rangle\langle \psi_j|\phi\rangle/|d_{\mathrm{max}}|^2,1\}$.
As  proved in Ref. \cite{observe}, $\mathcal{R}(D)$ is a convex function. Thus,
\begin{eqnarray}
C_f(\varrho)&=&\sum_j p_j C_f(|\psi_j\rangle)\geq\sum_j p_j \mathcal{R}(D^{(j)})\nonumber\\
&\geq&\mathcal{R}(\sum_j p_j D^{(j)})=\mathcal{R}(D),
\end{eqnarray}
where $D=\max\{\langle\phi|\varrho|\phi\rangle/|d_{\mathrm{max}}|^2,1\}$.

\end{document}